\documentclass[twocolumn,amsmath,amssymb,prb,showpacs]{revtex4}
\usepackage{graphicx}
\usepackage{amssymb}

\begin{document}

\title{Calculation of magnetic exchange couplings in S=3/2 honeycomb system Bi$_3$Mn$_4$O$_{12}$(NO$_3$) from first principles}

\author{Hem C. Kandpal}
\author{Jeroen van den Brink} 
\address{Institute for Theoretical Solid State Physics, IFW Dresden, D01171 Dresden, Germany}

\begin{abstract}
Absence of magnetic ordering in Bi$_3$Mn$_4$O$_{12}$(NO$_3$), (BMNO) which has a magnetic subsystem that consists of honeycomb bi-layers of  Mn$^{4+}$ ions with spin S=3/2, has raised the expectation that its ground state is strongly frustrated due to longer-range antiferromagnetic interactions. We calculate the magnetic exchange coupling constants of the BMNO complex within a density functional approach and find that the dominating interactions between Mn spins are the antiferromagnetic nearest-neighbor $J_1$ and interlayer interaction $J_c$. The largest interaction is $J_c$, which substantially exceeds $J_1$. Longer-range interactions are antiferromagnetic, but not strongly frustrating.
%only weakly frustrating.
\end{abstract}
\pacs{}

\date{\today}

\maketitle
In frustrated magnets spins interact through competing exchange interactions that cannot be simultaneously satisfied, giving rise to a large degeneracy of the system's ground state. The resulting fluctuations of the spins are at the root of remarkable collective phenomena such as emergent gauge fields and fractional particle excitations~\cite{Balents10}. 
In this context the particular magnetic properties of Bi$_3$Mn$_4$O$_{12}$(NO$_3$) (BMNO) have recently received considerable attention~\cite{SAK09,OEZ10,OKO10,MAT10,MGC10,GSK10}. BMNO is a layered manganese oxide with as its main magnetic building block a honeycomb lattice of Mn$^{4+}$ ions, each carrying a spin S=3/2. These manganese honeycomb lattices form bilayers (see Fig. 1), with interactions between the spins that are antiferromagnetic. In magnetic susceptibility and specific heat studies the absence of long-range ordering down to 0.4 K was reported~\cite{SAK09}, which was confirmed in a high-field ESR study~\cite{OEZ10}. It was subsequently observed by inelastic neutron scattering~\cite{MAT10} that long-range spin order can be induced by an external magnetic field. 

\begin{figure}
\includegraphics[width=0.35\textwidth,angle=0]{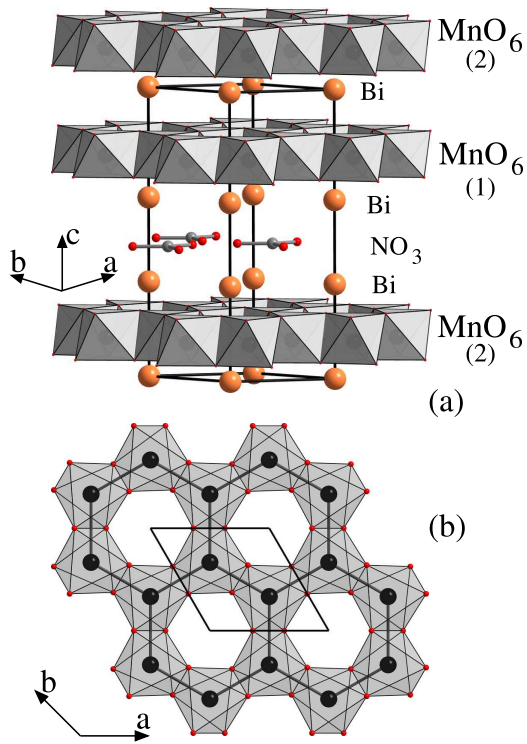}
\caption{(color online). Crystal structure of BMNO. (a). Distorted MnO$_6$ octahedra are edge sharing with oxygen  at the vertices, forming hexagonal bi-layers. For clarity the bottom hexagonal layer is repeated along $c$ direction, which emphasizes the bi-layer structure. (b). Top view of the network of edge-sharing octahedra in BMNO. The honeycomb lattice is indicated by the dark black lines that connect Mn atoms (black dots). The oxygen atoms shared by the MnO$_6$ units form layers. 
The orange (medium gray) spheres are Bi atoms, the yellow (light gray) spheres nitrogen atoms, and red (dark gray) spheres are oxygen atoms.%
}
\label{crystal}
\end{figure}

The lack of spontaneous magnetic ordering within the Mn honeycomb lattices down to very low temperature was proposed to be due to frustration of magnetic interactions, in particular a competition between first and second neighbor  antiferromagnetic Mn-Mn exchange interactions~\cite{SAK09}. Currently BMNO is therefore viewed as a model system with strongly fluctuating S=3/2 spins on honeycomb lattice, with a magnetic frustration that is due to the presence of longer-range antiferromagnetic exchange interactions~\cite{SAK09,OEZ10,OKO10,MAT10,MGC10,GSK10,WKW11}.

The usefulness of this paradigm critically depends on the exact values of the exchange magnetic interactions in this material. The exchange coupling constants have been extracted from inelastic neutron scattering experiments~\cite{MAT10}, where however the values of the couplings are not fully constrained. Theoretically, coupling constants have been estimated from unrestricted Hartree-Fock calculations and a perturbation method~\cite{WKW11}, which puts forward that a {\it ferromagnetic} third nearest neighbor interaction is responsible for the absence of long range magnetic order in BMNO.
We performed {\it ab initio} calculations of the magnetic exchange coupling constants of BMNO based on density functional theory and find that the nearest-neighbor $J_1$ and interlayer interaction $J_c$ are the most important interactions. In-plane longer-range interactions, including the third neighbor one, are antiferromagnetic and much smaller. 
Taking all these into account we find the system is significantly away from full frustration within the honeycomb planes.
%Magnetic frustration is therefore weak within the honeycomb planes. 
In contrast to what has been asserted so far for BMNO, we find that the dominant interaction is the coupling between the honeycomb bilayers, $J_c$, with a value that can be as large as twice $J_1$. The existence of interlayer interaction was recognized in a recent study by Ganesh \textit{et al.} in Ref.~[\onlinecite{GSK10}], but our finding that it actually dominates the exchange requires an alternate perspective on the factual origin of the observed magnetic frustration in BMNO.

\begin{figure}
\includegraphics[width=0.25\textwidth,angle=0]{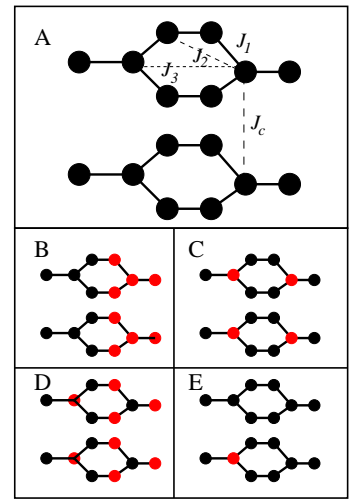}
\caption{(color online). Schematic representation of five ordered spin states A, B,
  C, D, and E in a 2$\times$2$\times$1 supercell of BMNO used for the determination of exchange
  constants. Only Mn atoms with the spin exchange paths are shown.
  Red filled (dark gray) circles are used to highlight spins that are flipped with respect to the
  reference configuration. Three in-plane magnetic interactions given as $J_1$, 
  $J_2$, and $J_3$, whereas $J_c$ represents the inter-bilayer interaction. }
\label{spin}
\end{figure}
 
We extract the exchange coupling constants in BMNO within GGA+U for different values of the Coulomb repulsion $U$ on the Mn site. So far BMNO has not been studied with first-principles calculations and we use frozen collinear spin configurations to calculate very accurately the energies for different spin configurations. For the calculations we used an all-electron full potential local-orbital scheme implemented in FPLO code~\cite{KEs99,fplo} with basis sets: Bi (5s, 5p, 5d, 6s, 6d, 6p, 7s, 7p), Mn (3s, 3p, 3d, 4s, 4p, 4d, 5s), N (1s, 2s, 2p, 3s, 3p, 3d), and O (1s, 2s, 2p, 3s, 3p, 3d) states.  The exchange-correlation energy functional was evaluated within the generalized gradient approximation (GGA) using the Perdew, Burke, and Ernzerhof parameterization~\cite{PBE96}. The number of irreducible \textit{k} points used in the calculations were 48 from a sampling of 4$\times$4$\times$3 mesh.

We use experimental crystal structure of Bi$_3$Mn$_4$O$_{12}$(NO$_3$), which is trigonal with space group $P3$ and consists of two MnO$_6$ layers separated by Bi and NO$_3$ layers (see figure~\ref{crystal})~\cite{SAK09}. The edge-shared MnO$_6$ octahedra lie in $ab$ plane and form honeycomb lattice as shown in figure~\ref{crystal}(b). 
The four inequivalent Mn atoms form two hexagonal planes separated by Bi atoms. These hexagonal planes are in distorted environments of MnO$_6$ octahedra resulting in small differences between the planes. We simplify the situation by treating the two layers as if they were the same, thus having the same in-plane magnetic interactions. The main purpose of presenting this streamlined picture is to understand and illustrate nature and main trends of magnetic interactions. But the small crystallographic differences will cause changes in the details of magnetic interactions that are presented here.
Five different magnetic spin configurations (as shown in Fig.~\ref{spin}) are considered to calculate the magnetic interactions within the honeycomb lattice and between the bilayers with the GGA+U method for $U$ = 4.0, 6.0, and 8.0 eV at the transition metal Mn site, which is within the range that of $U$ values extracted from spectroscopic measurements (4-5 eV)~\cite{Park96,Chinani93} and used in other {\it ab initio} calculations (4-10 eV)~\cite{Geng08,Satpathy96,Giovannetti08,Giovannetti09}. The Hund's rule coupling $J_H$ is 1 eV. 

Total energies for different Mn spin configurations in a supercell with 92 atoms were obtained in the GGA+$U$ formalism employing the atomic limit functional. The calculated relative energies for the spin configurations A, B, C, D, and E with the different $U$ values are listed in Table~\ref{tab1}, where the energy of the most stable configuration "D" energy was used as the zero of energy. With the energies of five ordered spin states, all four exchange constants ($J_1$,  $J_2$, and $J_3$, between first, second and third nearest neighbors, respectively and $J_c$ the inter-bilayer coupling) can be determined. 

\begin{table}
\begin{ruledtabular}
\begin{tabular}{l | r | r | r }
 & $U$=4.0 eV  & $U$=6.0 eV & $U$=8.0 eV   \\
\hline
A    & 0.3745   & 0.2213    & 0.1118 \\
B    & 0.1180   & 0.0684    & 0.0335 \\
C    & 0.1687   & 0.1013    & 0.0518 \\
D    & 0        & 0	    & 0      \\
E    & 0.2983   & 0.1752    & 0.0852 \\
\end{tabular}
\end{ruledtabular}
\caption{Relative energies in eV of different spin-states structures (as shown in Fig.~\ref{spin}) of BMNO obtained from GGA+U ($U$ = 4.0, 6.0, and 8.0 eV) calculations.}
\label{tab1}
\end{table}

To evaluate the magnetic interactions, we map the DFT energy differences of five magnetic spin configurations onto a classical spin-3/2 Heisenberg model. The magnetic exchange coupling constants are calculated from the following equations: 
%
%%%%%%%%%%%%%%%%%Begin equations %%%%%%%%%%%%%%%%%
\begin{align}
 E_A-E_B &= 9 \big(6J_1 + 12J_2 + 12J_3 \big)
\label{eq:EAB}\\
 E_A-E_C &= 9 \big(6J_1 + 12J_2 \big)
\label{eq:EAC}\\
 E_A-E_D &= 9 \big(12J_1 + 12J_3 \big)
\label{eq:EAD}\\
 E_A-E_E &= \frac{9}{2}\big(3J_1 + 6J_2 + 3J_3 + J_c \big)
\label{eq:EAE}
\end{align}
%%%%%%%%%%%%%%%%%End equations %%%%%%%%%%%%%%%%%
%
In Table~\ref{tab2}   $J_1$, $J_2$, $J_3$, and $J_c$ are listed as calculated from the relative energies of these spin configurations, for different values of $U$. All interactions come out to be antiferromagnetic and we observe that the dependence on $U$ is rather moderate as far as the relative values of the exchange constants is concerned: $U$ roughly tends to scale the different exchange interactions in the same manner. The in-plane $J_1$ is almost an order of magnitude larger than $J_2$. It is also considerably larger than $J_3$, so that these interactions are only weakly frustrating.  The magnitude of the couplings $J_2$ and $J_3$ are similar. As the bond length corresponding to $J_1$ (2.8697 \AA) is much shorter than that for $J_2$ or $J_3$ (4.9692 \AA/ or 5.7383 \AA), it is not surprising that $J_1$ is the strongest interaction within the honeycomb plane. However, the largest exchange interaction is the interlayer $J_c$, which is also antiferromagnetic and has about two times the value of $J_1$ for larger values of $U$.

\begin{table}
\begin{ruledtabular}
\begin{tabular}{l| c c c c}
exchange      & $U$ = 4.0 eV    & $U$ = 6.0 eV    & $U$ = 8.0 eV  \\ 	\hline
$J_1$	      & 3.00 (34.8)   & 1.7 (19.7)    & 0.9 (10.4)   	\\
$J_2$	      & 0.4 (4.6)     & 0.2 (2.3)     & 0.1 (1.2)	    	\\
$J_3$	      & 0.47 (5.4)    & 0.3 (2.7)     & 0.2 (2.3)	    	\\
$J_c$	      & 4.10 (47.6)   & 2.7 (31.3)    & 2.1 (24.4)    	\\
\hline
$J_2$/$J_1$   & 0.13333       & 0.1176        & 0.1111 	  	\\ 		   
$J_3$/$J_1$   & 0.15667       & 0.17647       & 0.2222 	      	\\     
\hline
$J_c$  (direct) &  2.97     &               1.88     &               1.08
\end{tabular}
\end{ruledtabular}
\caption{Exchange constants in BMNO in meV calculated by FPLO GGA+U with the atomic limit (AL) functional, on the basis of Eqs.~\ref{eq:EAB}-\ref{eq:EAE}. Slater parameters are chosen as $F_0=U$, $F_2=$8.6 eV and $F_4=$ 5.4 eV, i.e. $J=(F2+F4)/14=$ 1 eV. Values in Kelvin (K) in parentheses. The direct evaluation of $J_c$ is based on the energy difference between ferro and antiferro coupled FM honeycomb planes.
 }
\label{tab2}
\end{table}

Let us compare these results with the exchange constants found by inelastic neutron scattering. Matsuda \textit{et al.} have investigated magnetic ordered and disordered ground state using inelastic neutron scattering experiments~\cite{MAT10}. From the data they extract a constraint that relates the spin-spin interactions based to the observed magnetic excitations: 3$J_1$ + 6$J_2$ + $J_c$ $\sim$ 6 meV. 
%
%By subsequently using in addition the critical values of the coupling constants ($J_2/J_1$ = 0.15 and $J_c$/$J_1$ = 1/2) from references~\cite{Tak06,KIM86}, 
By subsequently using in addition an approximated value of $J_c$/$J_1$ = 1/2 from their experiments and the critical value of $J_2/J_1$ = 0.15 from literature~\cite{Tak06}, 
which ensure very strong frustration and are consistent with the observed absence of long-range magnetic ordering, Matsuda \textit{et al.} obtain $J_1$ = 2$J_c$ = 1.4 meV and $J_2$ = 0.20 meV. 
Compared to the experimental analysis, we consider in our theoretical calculations a fourth interaction ($J_3$) corresponding to the third nearest neighbors interaction within the honeycomb lattice. In this case the experimental constraint of  Matsuda \textit{et al.} is C=3$J_1$ + 6$J_2$ + $J_c$ + 3$J_3$ $\sim$ 6 meV, where we find from our calculations C=5.9 meV, for $U$ = 8.0 eV. The calculated constraint and therefore also the calculated exchange interaction are in full agreement with the dispersion of magnon excitations that were measured by inelastic neutron scattering.  We obtain a ratio of  $J_1$/$J_2$ that is similar to the one that was used by Matsuda \textit{et al.}~\cite{MAT10}. However, our interlayer coupling $J_c$ is larger than the in-plane coupling constants, 
%so that rather $J_c$/$J_1$ $\sim$ 2
contrary what was 
calculated from inelastic neutron scattering experiments~\cite{MAT10}. To cross-check this result, we also calculated the energies of fully ferromagnetically polarized honeycomb planes that are coupled either ferro- or antiferromagnetically. This allows for the direct evaluation of the intra-plane exchange but has the disadvantage that because a state with ferromagetic planes is the one furthest away from the in-plane antiferromagnetic ground state, significant renomalizations of the inter-plane magnetic bonds can occur. As shown in Table~\ref{tab2}, we find in this case a ratio of $J_c/J_1$ closer to one, the exact value of which depends on $U$. This is a substantial renormalization, but does not change the conclusion that the inter-layer coupling is the leading exchange interaction. This relatively large $J_c$ is related to the fact that the in-plane Mn-O-Mn angles are close to $90^0$, implying a strong competition between antiferromagnetic and ferromagnetic superexchange, of which it is a priori not clear which will win. The inter-plane exchange paths, via the covalent bismuth states, do not suffer from such a geometric frustration of interactions.

Thus from the theoretical evaluation of magnetic exchange coupling constants for BMNO, we find all of them (first-neighbors, second, and as well as third neighbors within the honeycomb plane and between the planes) to be antiferromagnetic. The calculated values fulfill the constraint that is put on the weighted sum of exchange interactions by inelastic neutron scattering data~\cite{MAT10}. In this constellation, only the second nearest neighbor interaction $J_2$ would be frustrating a full-blown antiferromagnetic ordering. As $J_2$ is comparable to $J_3$, both of which are much smaller than $J_1$, which is again exceeded by $J_c$, we find from the calculations that the effect of this frustration is not strong. 
We conclude that the observed absence of magnetic ordering down to temperatures lower than the temperature scale set by the calculated exchange interactions requires an explanation that goes beyond in-plane magnetic frustration. Possibly oxygen vacancies are of importance which will not only lead to disorder of magnetic bonds but might also lead to doping of the manganese layers, in which case the presence of Mn$^{3+}$ would lead to a competition between ferromagnetic double exchange and antiferomagnetic superexchange, which is known to strongly affect the magnetic properties and ordering in weakly doped  Mn$^{4+}$ perovskite manganites~\cite{Brink99}. At any rate, a description of the magnetic properties of BMNO will have to considered the dominating coupling constant, inter bi-layer exchange, explicitly.

We thank A. Paramekanti for stimulating our interest and discussing this honeycomb manganite with us and A. Tsirlin for discussions.

\end{document}